\documentclass[twocolumns,traditabstract]{aa}  
\usepackage{amssymb,amsmath}
\usepackage{changepage}
\usepackage{tabularx,multirow,booktabs,blindtext}
\usepackage{graphicx}
\usepackage{subfig}
\usepackage{placeins} 
\usepackage{txfonts}
\usepackage{graphicx,rotating}
\usepackage{subfig}
\usepackage{adjustbox}
\usepackage{txfonts}
\usepackage[T1]{fontenc}
\usepackage{epsfig}
\usepackage{natbib}
\usepackage{color}
\usepackage{tabularx}
\usepackage{comment}
\usepackage{longtable}
\usepackage{lipsum}
\usepackage{ltablex}
\usepackage{caption, booktabs}

\def\eg{e.g.}

\def\bggi{BGG 0957+0204}
\def\bggii{BGG 1000+0207}
\def\bggiii{BGG 0958+0223}

\DeclareGraphicsExtensions{.pdf,.png,.jpg}
\bibpunct{(}{)}{;}{a}{}{,}
\usepackage{hyperref}
\hypersetup{
    colorlinks=true,
    linkcolor=blue,
    citecolor=blue,
    filecolor=magenta,
    urlcolor=cyan
}

\defcitealias{kartaltepe_multiwavelength_2010}{Kartaltepe+10}
\defcitealias{weaver_cosmos2020_2022}{COSMOS2020}

\begin{document} 

\authorrunning{Toni et al.}
\titlerunning{Molecular gas properties of star-forming brightest group galaxies at $z \sim0.3$}

\title{Molecular gas properties of star-forming \\brightest group galaxies at $z \sim0.3$}

   \author{
   Greta~Toni\inst{1,2,3,}\thanks{Corresponding author: \texttt{greta.toni4@unibo.it}}
   \and   
   Gianluca~Castignani\inst{2}
   \and
   Françoise~Combes\inst{4}
   \and
   Philippe~Salomé\inst{4} 
   \and 
   Angel~Bongiovanni\inst{5} 
   \and
   Lauro~Moscardini\inst{1,2,6} 
   \and 
   Matteo~Maturi\inst{3,7}}
   \institute{
   Dipartimento di Fisica e Astronomia “A. Righi”, Alma Mater Studiorum Università di Bologna, via Gobetti 93/2, 40129 Bologna, Italy
   \and 
   INAF - Osservatorio  di  Astrofisica  e  Scienza  dello  Spazio di  Bologna,  via  Gobetti  93/3, 40129  Bologna,  Italy 
   \and
    Zentrum für Astronomie, Universität Heidelberg, Philosophenweg 12, 69120 Heidelberg, Germany
    \and
Observatoire de Paris, Collège de France, PSL University, Sorbonne University, CNRS, LUX, 75014 Paris, France
\and
Institut de Radioastronomie Millimétrique (IRAM), Av. Divina Pastora 7, Núcleo Central 18012, Granada, Spain
\and
INFN - Sezione di Bologna, viale Berti-Pichat 6/2, 40127 Bologna, Italy
\and ITP, Universität Heidelberg, Philosophenweg 16, 69120 Heidelberg, Germany}

  \abstract
  {Recent efforts to characterise the molecular gas content of brightest cluster galaxies (BCGs) at intermediate redshift have revealed a sub-population of gas-rich systems, whose star formation activity is likely influenced by environmental processing. In this study, we aim to investigate the molecular gas reservoirs and star formation fuelling of central galaxies in groups, also known as brightest group galaxies (BGGs), at intermediate redshifts. We present targeted carbon monoxide (CO) line observations of three BGGs in the COSMOS field at $z \sim 0.3$, obtained with the IRAM 30m telescope. The galaxies exhibit disturbed morphologies, extended blue substructures, and interaction signatures. Furthermore, they exhibit significant star formation rates derived from multiwavelength diagnostics. We detect CO(1$\rightarrow$0) emission in one system, revealing a substantial molecular gas mass of $M_{H_2} \sim 3 \times 10^{10}$~M$_\odot$, while for the other two BGGs, CO emission lines remain undetected, yielding stringent upper limits of $M_{H_2} \lesssim 10^{10}$~M$_\odot$. By combining molecular gas constraints with fiducial star formation rates derived from total infrared emission, we infer gas depletion timescales in the range of $\lesssim0.5-1.5$~Gyr. These results may indicate that, despite their active star formation and interaction signatures, some BGGs could already experience efficient gas exhaustion or suppressed gas replenishment, suggesting that gas depletion precedes star formation quenching. Our findings hint that environmental processes in galaxy groups could strongly regulate the availability of cold gas and drive rapid evolutionary phases in central galaxies, possibly bridging the gap between gas-rich BCGs and passively evolving systems.}


   \maketitle

\section{Introduction}

Brightest cluster galaxies (BCGs) and brightest group galaxies (BGGs) are remarkable laboratories for exploring the effect of dense environments on the evolution of galaxies. Their large masses and luminosities are believed to be the aftermath of a complex and still-debated evolutionary history. In clusters, BCGs have been extensively studied to understand the build-up of their stellar mass, their quenching pathways, and their links to the host halo properties and the intracluster light \citep[\eg][]{de_lucia_hierarchical_2007, lin_k-band_2004, kravtsov_stellar_2018}. Currently favoured paradigms propose that while the bulk of their stars are formed at early stages ($z>3$), a consistent assemble of mass takes place via minor dry mergers at lower redshift, significantly increasing the stellar content of the galaxy, although with potentially strong dependence on the host halo and large-scale environment \citep[\eg][]{de_lucia_hierarchical_2007, stott_little_2011, lidman_evidence_2012, einasto_galaxy_2024}.
Observations and simulations consistently indicate that the assembly histories of BCGs lead to the formation of massive elliptical galaxies, which consequently evolve passively, showing quenched star formation, at least in the local Universe \citep[\eg][]{martizzi_formation_2012,zhao_exploring_2017,chu_physical_2022}. For instance, \citet{chu_physical_2022} analysed a large sample of more than a thousand BCGs, finding that only $\sim$7--9\% exhibit indications of residual star formation at $z < 0.7$.

The baryonic budget of BCGs—including their molecular gas, dust, and ionised components—is closely linked to their stellar assembly and overall evolutionary pathways. Numerous studies in the local Universe have revealed that some BCGs host substantial reservoirs of molecular gas \citep{edge_detection_2001, salome_cold_2003, hamer_relation_2012, mcnamara_1010_2014, russell_bow_2014}, and in several cases extended cold or ionised filaments \citep{olivares_ubiquitous_2019, russell_driving_2019}. In cool-core clusters, molecular gas masses of the order of $10^{10}$--$10^{11}\,\rm M_\odot$ are common \citep[\eg][]{fogarty_dust_2019, castignani_molecular_2025}, often co-spatial with ionised and dusty filaments \citep[\eg][]{olivares_ubiquitous_2019}. Such configurations likely trace intracluster medium (ICM) condensation, which can fuel residual star formation or active galactic nuclei (AGN) activity \citep[\eg][]{fogarty_dust_2019, castignani_molecular_2025, ubertosi_jvla_2025}. 

Although rare at higher redshift, some BCGs were found to exhibit gas-rich phases, such as the Phoenix~A BCG, which hosts a $\sim2\times10^{10}\,\rm M_\odot$ molecular reservoir at $z=0.597$ \citep{mcdonald_state_2014} with an extreme star-formation rate (SFR) of $\sim800 \, \rm M_\odot \,\mathrm{yr}^{-1}$ \citep{mcdonald_hstwfc3-uvis_2013}, implying rapid assembly of $\sim$3\% of its stellar mass within $\sim$30 Myr.

Compared to BCGs, BGGs, the central galaxies of galaxy groups, have historically received less attention despite their importance: groups are the most common environment for galaxies, and baryonic processes such as cooling, AGN feedback, and mergers can play a dominant role in shaping their baryonic cycle as well as the properties of their low-velocity-dispersion haloes. Recent work shows that BGGs do not simply scale down BCG evolution but instead follow more diverse and sometimes distinct evolutionary pathways \citep[see \eg][]{2026arXiv260210192B}.
In the COSMOS X-ray group sample, \citet{gozaliasl_chandra_2019} find that $\sim20\%$ of local BGGs are star-forming, increasing to $\sim50\%$ at $z\sim1$. Their stellar ages are also systematically younger than predicted by semi-analytic models \citep{gozaliasl_cosmos_2024}, especially for systems offset from X-ray centres, possibly signalling recent group mergers \citep{george_galaxies_2011, gozaliasl_chandra_2019}. Earlier results from the same series \citep{gozaliasl_brightest_2018} show a bimodality between quiescent and star-forming BGGs, with the latter exhibiting mean SFRs $\sim$2 dex above the overall median SFR and significant in situ stellar mass growth. A non-negligible fraction of BGGs at $z\lesssim$ 0.3 still form stars \citep{gozaliasl_cosmos_2024}, suggesting that in some cases residual or accreted cold gas—potentially condensing from the intragroup medium (IGM; e.g. \citealt{temi_alma_2018, olivares_gas_2022})—continues to fuel star formation or AGN activity in group cores. Finally, recent work further emphasises the diversity of central galaxy growth within groups, spanning star formation histories, kinematics, and structural properties \citep[\eg][]{jung_massive_2022}, and shows that BGG properties closely correlate with those of their host haloes and even with their Mpc-scale environment \citep[\eg][]{raouf_sami_2021, einasto_galaxy_2024}.
This highlights the importance of group-scale haloes in understanding the balance between cooling, feedback, and star formation in relation to the environment. In this context, a detailed investigation of BGG formation and evolution is both timely and compelling. 

Carbon monoxide (CO) emission is the primary tracer of molecular hydrogen (H$_2$), whose detection is challenging due to the lack of a permanent dipole moment. CO, by contrast, is abundant, has low excitation temperature ($T_{\mathrm{ex}} \sim 5$~K for the CO(1$\rightarrow$0) transition), and its rotational transitions fall conveniently in the millimetre/submillimetre regime, where they can be more easily observed. Thus, CO emission provides a reliable proxy for mapping the distribution, kinematics, and total mass of the molecular gas, given an assumed CO-to-H$_2$ conversion factor ($\alpha_{\mathrm{CO}}$). Altogether, detecting CO line emission in BCGs/BGGs provides a direct probe of the gas that can fuel star formation and AGN activity \citep[\eg][]{edge_detection_2001,salome_cold_2003}.  Conversely, non-detections of CO can indicate effective and rapid quenching, providing stringent tests of theoretical models of galaxy–host coevolution.

This work aims to explore the environmental processing of three star-forming central galaxies in COSMOS groups at intermediate redshifts ($z \sim 0.3$) by characterising their molecular gas reservoirs through targeted emission line observations with the IRAM 30m telescope (ID 079-25, PIs: Toni \& Castignani).
Namely, we targeted three star-forming BGGs to confirm (or constrain, in case of non-detections) their molecular gas reservoirs. These observations extend previous CO campaigns of intermediate- and high-redshift BCGs with the IRAM 30m \citep{castignani_molecular_2019, castignani_molecular_2020,Castignani_clash_newitem_2020, castignani_star-forming_2022, castignani_star-forming_2023} into the group regime. This pilot program forms the basis of an observation campaign which will ultimately enable the construction of the first statistically significant sample of group and cluster central galaxies at intermediate-to-high redshift with robust CO-observation-based characterisation of molecular gas properties.

Our program is designed to constrain the molecular gas content and star formation efficiency of central galaxies in group environments, providing insight into gas processing at $z \sim $ 0.3 (lookback time $\sim 3$~Gyr). Despite growing efforts, current literature still lacks sizeable samples of distant central galaxies in dense environments with molecular gas observations feeding the star formation \citep[see \eg][]{dunne_co_2021}. The lack of such observations is even more significant when we move to the group regime \citep[\eg][]{olivares_gas_2022}, which traces intermediate environments, having intermediate galaxy densities, with values in between those of the field and the dense clusters. By comparing the derived molecular gas masses ($M_{H_2}$), gas-to-stellar mass fractions, depletion times, excitation conditions, and SFRs with those of cluster and field galaxies across cosmic time \citep[\eg][]{tacconi_phibss_2018, freundlich_phibss2_2019, fogarty_dust_2019, dunne_co_2021}, we aim to assess the role of group-scale environments in regulating gas accretion, star-formation fuelling, and potential rapid gas exhaustion.

This paper is structured as follows. In Sect.~\ref{sec:BGG_sample}  we report a description of the three star-forming BGGs analysed in this study, including the target selection and the SFR estimation through different diagnostics. In Sect.~\ref{sec:mm_observations}, we describe our millimetre observations of the targets with the IRAM 30m telescope and the relative data reduction. Section \ref{sec:discussion} is dedicated to the discussion of the results and the emerging picture about the star formation activity and gas consumption in the three BGGs. Finally, Sect.~\ref{sec:conclusions} summarises the main results of this work and outlines future perspectives for this observation campaign.
Throughout this work, we adopt a flat $\Lambda \rm CDM$ cosmology with matter density $\Omega_{\rm m} = 0.3$, dark energy density $\Omega_{\Lambda} = 0.7$ and Hubble constant $h=H_0/100\, \rm km\,s^{-1}\,Mpc^{-1} = 0.7$.

\section{Star-forming brightest group galaxies} \label{sec:BGG_sample}

We selected our targets among the central galaxies of the clusters and groups of the AMICO-COSMOS catalogue \citep{toni_amico-cosmos_2024}, detected with the AMICO algorithm \citep{bellagamba_amico_2018,maturi_amico_2019} in the COSMOS field \citep{scoville_cosmic_2007}, using the COSMOS2020 \citep{weaver_cosmos2020_2022} and the COSMOS2015 \citep{laigle_cosmos2015_2016} galaxy catalogues. The adopted cluster and group catalogue covers an effective area of $\sim 1.69$ deg$^2$. Despite the relatively small area that characterises the COSMOS field, this catalogue yields a sufficiently large and statistically robust sample of BGGs in the redshift range considered in this study. The AMICO-COSMOS catalogue provides information about the cluster or group detections, including sky position, redshift, richness, X-ray luminosity, and total mass. Alongside the detection catalogue, each identified system is accompanied by a list of associated member galaxies with their membership probabilities. For details about the cluster and group catalogue with members, we refer the reader to \citet{toni_amico-cosmos_2024}, where the AMICO-COSMOS catalogue is presented.

Based on their richness and the mass estimates from \citet{toni_amico-cosmos_2024} spanning $M_{200} \sim 6 \times 10^{12}$–$3 \times 10^{14}\, \rm M_\odot$\footnote{Normally used as a reference mass for haloes, $M_{200}$ is the mass defined with respect to $R_{200}$, the radius of the sphere in which the mean density is 200 times the critical density of the Universe.}, the systems we analyse are predominantly galaxy groups. Recent studies of the COSMOS field show that only a few systems reach cluster-scale masses ($>10^{14}\, \rm M_\odot$) or have more than 50 members, which are commonly adopted criteria to distinguish groups from clusters \citep[\eg][]{knobel_zcosmos_2012,gozaliasl_chandra_2019,toni_amico-cosmos_2024,toni_cosmos-web_2025}. Therefore, we refer to our systems simply as galaxy groups and to their central galaxies as BGGs.

\begin{table*}[]
\caption{Properties of our selected targets: three star-forming BGGs in COSMOS.}
\centering
\begin{center}
\begin{tabular}{ccccccccc}
\hline\hline\\
Galaxy & ID & R.A. & Dec. & $z_{spec}$ &  $M_\star$ & SFR & sSFR & sSFR$_{\rm MS}$  \\
   & COSMOS2020 & [hh:mm:ss.s] & [dd:mm:ss.s] &  &  [$10^{10}$~M$_\odot$] & [M$_\odot$~yr$^{-1}$] & [Gyr$^{-1}$] & [Gyr$^{-1}$]  \\ 
  (1) & (2) & (3) & (4) & (5) & (6) & (7) & (8) & (9) \\\\
 \hline  \\
BGG~0957+0204 & 683413& 09:57:47.570 & 02:04:15.880      & 0.25288 & $1.2^{+0.1}_{-0.1}$  &  $6.7^{+1.1}_{-1.1}$  & $0.55^{+0.11}_{-0.11}$  &    $0.16$   
 \\\\
\hline  \\
BGG~1000+0207 & 745502 & 10:00:10.940 &  02:07:24.760      & 0.33904 &  $6.3^{+1.1}_{-1.0}$  & $29.8^{+3.6}_{-3.6}$  & $0.48^{+0.10}_{-0.10}$  & 0.11   
 \\\\
\hline  \\
\\
 BGG~0958+0223 & 1041562 & 09:58:38.830 & 02:23:48.750 &     0.35510 &  $6.9^{+0.8}_{-0.7}$  & $45.8^{+5.1}_{-5.1}$  & $0.67^{+0.11}_{-0.10}$  &   0.11   
 \\\\
\hline

\end{tabular}
 \end{center}
 \tablefoot{Column description: (1) BGG name; (2) unique identifier in COSMOS2020 \citep{weaver_cosmos2020_2022}; (3-4) J2000 equatorial coordinates; (5) spectroscopic redshift\textsuperscript{\ref{fn:desi}}; (6) stellar mass \citep{weaver_cosmos2020_2022}; (7)  SFR derived from spectral energy distribution modelling of FIR emission  and using the \citet{1998ARA&A..36..189K} relation, which we take as fiducial value  (see Sect. \ref{sec:SFR}); (8) specific SFR, i.e. ${\rm sSFR}={\rm SFR}/M_\star$; (9) sSFR for main sequence field galaxies with redshift and stellar mass of our targets estimated using the relation found by \citet{speagle_highly_2014}.}
\label{tab:galaxy_properties}
\end{table*}

\subsection{BGG catalogue}
Identifying BCGs/BGGs in photometric cluster and group catalogues is a non-trivial task, since no single observable provides a fully reliable tracer of the central galaxy. In the AMICO framework, membership probabilities depend not only on projected distance from the group centre, but also on galaxy luminosity and on the shape of the photometric-redshift probability distribution. Although these observables are all considered within the AMICO detection and membership assignment procedure, the membership probability alone is not always a reliable tracer of BGGs. In particular, the assigned membership probability is sensitive to the shape and peakiness of the galaxy photometric-redshift probability distribution \citep[as previously shown by \eg][]{toni_amico-cosmos_2024,euclid_collaboration_bhargava_euclid_2025}, which is modelled as a Gaussian distribution with mode and 1$\sigma$ errors taken from the photometric galaxy catalogue. Additionally, Spectral Energy Distribution (SED) fitting and therefore accurate photo-$z$ estimation can be challenging for bright extended galaxies at low and intermediate redshift, where galaxy morphology and substructures are spatially resolved \citep[\eg][]{soo_morpho-z_2018,wilson_photometric_2020,weaver_cosmos2020_2022}. Consistent with this effect, we identified several bright extended central galaxies in our sample (particularly at $z \lesssim0.5$) that clearly exhibit the expected properties of BGGs, but were assigned low AMICO membership probability because their photometric-redshift probability distributions were, for instance, sharply peaked at values slightly offset from the system redshift (i.e. from the average redshift of the other member galaxies). Although this effect impacts only a small fraction of systems, it can influence BGG selection, particularly within the redshift range relevant to this study. For this reason, we constructed the AMICO-COSMOS BGG catalogue using a selection strategy that primarily combines galaxy luminosity and proximity to the AMICO group centre, with the membership probability adopted only as a secondary decision criterion. Similar methodologies were previously validated in AMICO-based studies \citep[\eg][]{radovich_amico_2020, gozaliasl_cosmos_2025}. Therefore, we proceeded according to the following steps:
\begin{itemize}
    \item vicinity to the system centre (sky distance), $\Delta x$: we filtered the galaxy sample by keeping only galaxies with $\Delta x < 0.3 R_{200}$, where $R_{200}$ is the radius of the AMICO cluster model (which does not necessarily correspond to the $R_{200}$ of individual detections), the same used for the definition of the intrinsic richness, $\lambda_\star$ (see e.g. \citealt{maturi_amico_2019} and \citealt{toni_amico-cosmos_2024}), and 0.3 is a typical value used to limit the search to the central area of each candidate group;
    \item vicinity to the system redshift, $\Delta z$: a constraint in redshift is needed, since we do not use any a priori limitation in membership probability. To limit the selection of the BGG to the vicinity of the group candidate, also in redshift space, we imposed $|\Delta z| < 0.05(1+z)$, where $z$ is the system redshift. This choice is in line with the average photometric redshift probability uncertainty of COSMOS2015 and COSMOS2020 galaxies;
    \item galaxy magnitude, $m$: among the obtained galaxy sub-sample, we selected the most likely BGG as the galaxy with the lowest apparent magnitude $m$; 
    \item membership probability, $P$: only as a final step, if the first and the second brightest galaxies have a magnitude difference $<0.5$, we selected the one with the highest membership probability between the two.
    
\end{itemize} 

\begin{figure*}[h]
   \centering
   \includegraphics[width=\linewidth]{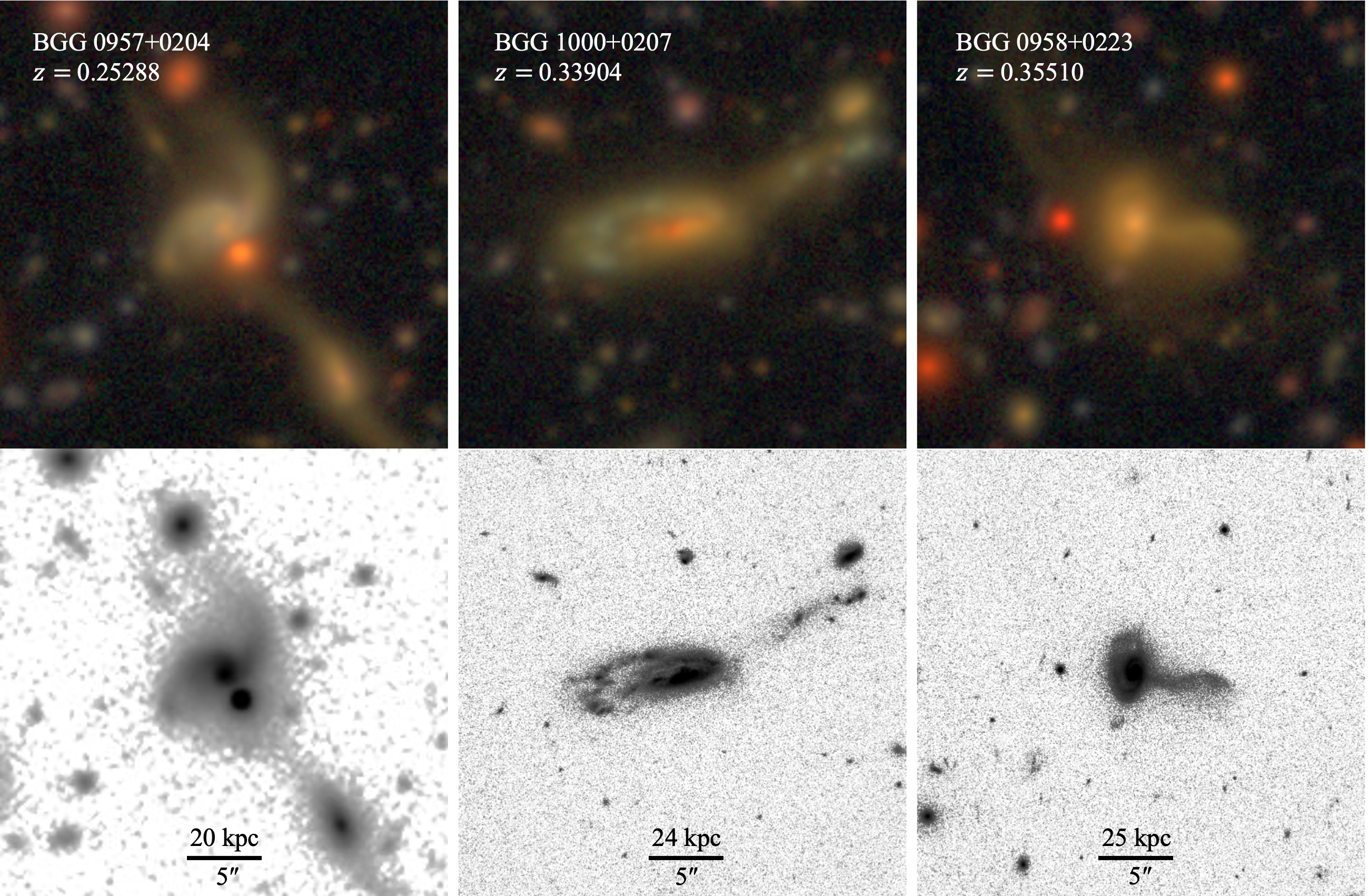}
      \caption[Colour composites of our three targets]{From left to right: Target BGG~0957+0204, BGG~1000+0207 and BGG~0958+0223. Top panels: Colour composite 30$^{\prime\prime}$$\times$30$^{\prime\prime}$ images of the three targets in HSC $g$, $r$, $i$ bands; bottom panels: 30$^{\prime\prime}$$\times$30$^{\prime\prime}$ UltraVISTA Ks image (for BGG~0957+0204) and HST-ACS I mosaic (for BGG~1000+0207 and BGG~0958+0223). For reference, at the bottom of each panel we report the physical size corresponding to 5$^{\prime\prime}$. Our targets are unresolved by our observations, as the IRAM~30m HPBW falls in the range $\sim (26 - 28)$$^{\prime\prime}$ and $\sim (8.7 - 9.4)$$^{\prime\prime}$ for CO(1$\rightarrow$0) and CO(3$\rightarrow$2), respectively (see Sect. \ref{sec:IRAM30m_observations} for details). North is up, east is left.}
         \label{fig:images}
\end{figure*}

\noindent For all the detections in the group catalogue, we stored the first and second choices according to these criteria. 
It should be noted that the AMICO-COSMOS catalogue is composed of three different catalogues resulting from three independent detection runs, each taking a different photometric band ($r$, $Y$, or $H$) as the reference band \citep{toni_amico-cosmos_2024}. We created a BGG catalogue for each of the runs using the magnitude in the same photometric band used for detection as the defining galaxy magnitude, $m$. The final catalogue of central galaxies is the result of the merging of the three catalogues, keeping any unique selected candidate, in a way that each system can have multiple options for its chosen BGG. As we have used the input photometric galaxy catalogue chosen for the AMICO-COSMOS group search, galaxies here selected and their photometric properties are taken from COSMOS2020 CLASSIC \citep{weaver_cosmos2020_2022} or, in a few cases \citep[see][for details]{toni_amico-cosmos_2024}, COSMOS2015 \citep{laigle_cosmos2015_2016}.

Each BGG in the catalogue is then assigned two scores, one related to its association probability and one related to its distance on the sky plane to the centre of the host system. The first score is the ratio of the BGG membership probability to the highest probability assigned by AMICO in that group. The second score is the distance to the centre in units of $R_{200}$, which is again the AMICO model radius. The combination of these two is an indicator of the probability of the galaxy being the BGG of the system. We have expanded and utilised this BGG selection method also for the group catalogue based on the COSMOS-Web data, presented by \citet{toni_cosmos-web_2025}, a more area-limited but much deeper group catalogue extending up to $z=3.7$. The resulting COSMOS-Web BGG catalogue has been exploited for the investigation of size--mass evolution of central group galaxies. The work is presented in \citet{gozaliasl_cosmos_2025}.

\subsection{Selection of targets}
As we aim to search for CO emission lines, we cross-matched the selected BGGs with high-confidence measurements from public spectroscopic redshift catalogues (the same listed in \citealt{toni_amico-cosmos_2024}, Sect. 5.4). In the case of aforementioned scores being low, we used spec-$z$s to rule out ambiguous cases. The spectroscopic counterpart assignment yielded $\sim 800$ sources with spec-$z$ consistent with the redshifts of the associated parent group. 
Among the galaxies with spectroscopic redshifts, we selected those at intermediate redshifts, namely in the range $0.2<z<0.4$. This redshift interval was chosen to target non-local galaxies without compromising the feasibility of the observations with the IRAM 30m telescope. Then, we further limited ourselves to star-forming sources, selecting only those with $\textrm{SFR}>20~\rm M_\odot$~yr$^{-1}$ according to COSMOS2020 physical properties derived via SED fitting with LePhare \citep{arnouts1999, ilbert_accurate_2006}. However, as it will be further discussed in Sect. \ref{sec:SFR}, we refined the star-formation-rate estimates a posteriori to improve the reliability of our analysis. The final refined estimates are those also reported in Table \ref{tab:galaxy_properties}.

The selection procedure yielded three central galaxies, all belonging to groups with masses in the range $M_{200}\simeq (1 - 3) \times10^{13}$~M$_\odot$, thus lower than those typically associated with galaxy clusters.
As reported in Table \ref{tab:galaxy_properties}, our targets have intermediate redshifts, $z\sim0.25-0.35$\footnote{\label{fn:desi}spectroscopic redshifts of these three targets are all DESI-EDR and DESI-DR1 redshifts \citep{adame23, desi_collaboration_data_2025}, also reported in the COSMOS Spectroscopic Redshift Compilation \citep{khostovan_cosmos_2025}.}, high stellar masses, $M_\star\simeq(1-7)\times10^{10}$~M$_\odot$, typical of ellipticals at the centre of clusters and groups, and are clearly star-forming (SFR multiwavelength diagnostics will be discussed below in Sect. \ref{sec:SFR}). These galaxies may therefore represent the intermediate-$z$ analogues of local star-forming central galaxies such as Perseus~A and Cygnus~A \citep[\eg][]{salome_very_2011,privon_modeling_2012,fraser-mckelvie_rarity_2014}. Table \ref{tab:galaxy_properties} summarises the main physical properties of the three targets, including the expected SFR for a galaxy of the same mass at the same redshift on the star-forming main sequence (MS) according to \citet{speagle_highly_2014}. Star formation rates reported in Table \ref{tab:galaxy_properties} are the fiducial values we assume for this study, which are extracted by integrating dust emission after performing SED fitting. Multiwavelength indicators and SFR estimates are described in Sect. \ref{sec:SFR}.

Our three selected targets are displayed in Fig.~\ref{fig:images}, where the upper panels show colour-composite Hyper Suprime Cam (HSC) images \citep{aihara_third_2022}, while the bottom panels show UltraVISTA Ks \citep{mccracken_ultravista_2012} and Hubble Space Telescope's ACS \citep{koekemoer_cosmos_2007} single-band images. The images display hints of spiral arms and a clearly disturbed morphology for the three targets, possibly induced by the interaction with the surrounding companions, suggesting strong processing of gas and influence of the environment. 
 \\

\begin{figure}[h]
   \centering
   \includegraphics[width=\linewidth]{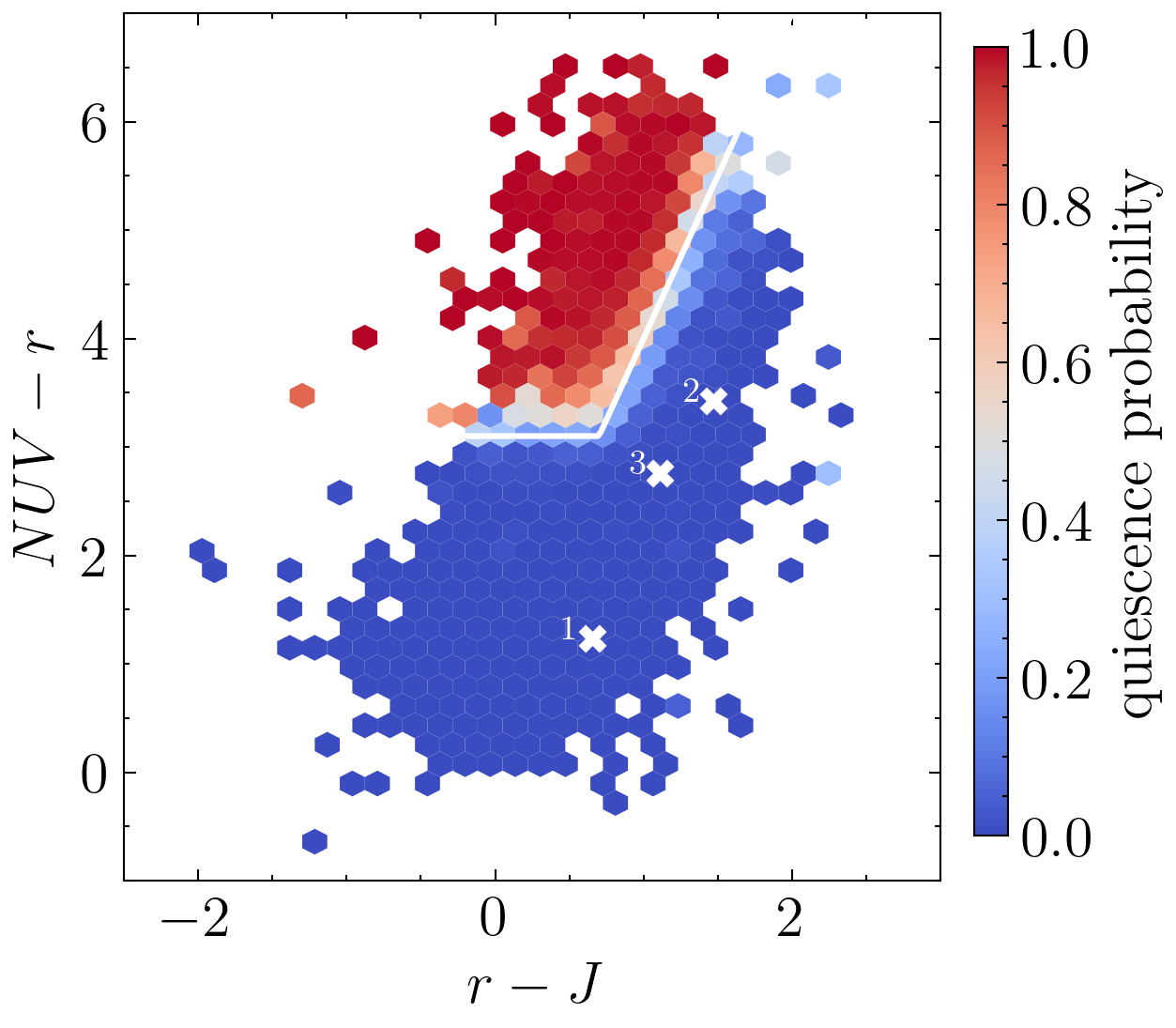}
      \caption{Location of our three targets (white crosses: 1. \bggi; 2. \bggii; 3. \bggiii) in the NUVrJ diagnostic diagram of the AMICO-COSMOS member galaxies. Colours indicate the mean probability to be quiescent in each hexagonal bin, as in the bar on the right. Rest-frame colours of the target galaxies are clearly separated from the passive population (red hexagons).}
         \label{fig:nuvrj}
\end{figure}

\subsection{Star formation rates}\label{sec:SFR}
Our three target BGGs have a probability of being star-forming larger than 99\% according to the machine-learning-based classification method developed by \citet{toni_redsequence_2025}. Rest-frame diagnostic diagrams like NUVrJ \citep{ilbert_mass_2013} further confirm that the sources are star-forming and clearly separate themselves from the population of quiescent galaxies. In Fig. \ref{fig:nuvrj}, we report the position of our three selected targets in the NUVrJ diagram of the AMICO-COSMOS member galaxies with quiescence probabilities from \citet{toni_redsequence_2025}.

A key component of our analysis is the connection between star-formation activity and properties of the molecular gas reservoirs, for which we need reliable and justified estimates of the SFR. The COSMOS2020 catalogue \citep{weaver_cosmos2020_2022} includes extensive multi-wavelength photometry from the ultraviolet through the near- and mid-infrared. Specifically, it combines deep imaging from GALEX, HSC, UltraVISTA, and Spitzer-IRAC over the wavelength range $\sim0.15-5\,\mu$m \citep{weaver_cosmos2020_2022}. However, this wavelength coverage does not include the full mid-infrared regime nor extend into the far-infrared (FIR; $\gtrsim 20~\mu$m) range, where the bulk of dust-reprocessed emission associated with star formation emerges, as typically found in massive and dusty galaxies \citep[\eg][]{kennicutt_star_2012, casey_are_2014}. The COSMOS2020 catalogue's SED fitting is solely based on the optical and near-infrared part of the spectrum, which often fails to fully and reliably trace obscured star formation in systems with significant dust content \citep[\eg][]{casey_are_2014, leja2017, malek2018}.

Independent tracers of star formation, such as emission lines (H$\alpha$ and [OII]) and total FIR luminosities, therefore play a crucial role in building a comprehensive view of star formation activity \citep[see \eg][for a review]{calzetti2013}. The emission lines provide information on recent star formation on $\sim$10 Myr timescales, while FIR luminosities trace the reprocessed light from young stars embedded in dust and longer timescales, of the order of $\sim$ 100 Myr \citep[\eg][]{calzetti2025}. By comparing SFRs obtained from different estimators, we can assess their agreement and interpret them in light of the timescales they probe. 
Table \ref{tab:sfrs} reports the results in terms of SFR and corresponding sSFR for the multiwavelength indicators we analyse in this Section.

\subsubsection{\textit{Line diagnostic from DESI spectra}}
\label{sec:SFR_DESIspectra}

We investigate the star formation properties of the BGGs using DESI DR1 spectroscopy \citep{desi_collaboration_data_2025}, following the general approach adopted in previous studies \citep[\eg][]{castignani_star-forming_2022, castignani_star-forming_2023}. Star-formation rates are estimated from emission-line diagnostics, exploiting the proportionality between the SFR and the luminosities of the H$\alpha$ recombination line and the [OII] forbidden-line doublet at 3726~\AA\ and 3729~\AA\ \citep{1998ARA&A..36..189K}. Throughout this work, all SFRs are calibrated to a \citet{Chabrier2003PASP..115..763C} initial mass function (IMF).

Emission-line fluxes are retrieved from the DESI DR1 Galaxy and Quasar/AGN Value-Added Catalogues \citep[VAC;][]{myers_target-selection_2023}, which provide measurements derived from the \textsc{FastSpecFit} spectral fitting pipeline \citep{fastspec}. These fluxes correspond to the narrow emission-line components measured after subtraction of the stellar continuum.

Because DESI spectra are obtained through fibres with a diameter of 1.5~arcsec, aperture corrections are required for extended galaxies. We correct all emission-line fluxes using the aperture correction parameter provided by the \textsc{FastSpecFit} Emission-Line Catalog\footnote{\url{https://data.desi.lbl.gov/doc/releases/dr1/vac/fastspecfit/}}, which estimates the ratio between the total galaxy flux and the flux captured within the fibre based on the best-fit morphological model from Legacy Survey imaging \citep{dey_overview_2019,desi_collaboration_data_2025}. When multiple spectral measurements are available for the same galaxy, we adopt the spectrum based on the largest number of coadded exposures, since this generally provides the highest signal-to-noise ratio and therefore the most robust redshift and emission-line measurements.

\begin{table}[]
\caption{SFRs and relative sSFR of our target BGGs.}
\centering
\begin{center}
\begin{tabular}{cccc}
\hline
\hline
\noalign{\smallskip}
star-formation & BGG & BGG & BGG \\
indicator & 0957+0204  & 1000+0207   & 0958+0223  \\
\noalign{\smallskip}
\hline 
\noalign{\smallskip}
\noalign{\smallskip}
FIR & $6.7^{+1.1}_{-1.1}$  & $29.8^{+3.6}_{-3.6}$ & $45.8^{+5.1}_{-5.1}$ \\
\noalign{\smallskip}
 $^\mathrm{(this \, work)}$ & $0.55^{+0.11}_{-0.11}$ & $0.48^{+0.10}_{-0.10}$ & $0.67^{+0.11}_{-0.10}$ \\
 \noalign{\smallskip}
 \noalign{\smallskip}
 \hline
 \noalign{\smallskip}
 \noalign{\smallskip}
 FIR & --  & $49.6^{+3.2}_{-3.2}$ & $30.3^{+2.8}_{-2.8}$ \\
\noalign{\smallskip}
 $^{\rm(\citetalias{kartaltepe_multiwavelength_2010})}$ & -- & $0.79^{+0.14}_{-0.15}$ & $0.44^{+0.06}_{-0.07}$ \\
 \noalign{\smallskip}
 \noalign{\smallskip}
 \hline
 \noalign{\smallskip}
 \noalign{\smallskip}
 H$\alpha$  & $13.3^{+1.3}_{-1.3}$  & $11.8^{+3.6}_{-3.6}$ & $61.9^{+9.4}_{-9.4}$ \\
 \noalign{\smallskip}
 & $1.11^{+0.15}_{-0.17}$ & $0.19^{+0.07}_{-0.07}$  & $0.90^{+0.16}_{-0.17}$ \\  \noalign{\smallskip}
 \noalign{\smallskip}
 \hline
 \noalign{\smallskip}
 \noalign{\smallskip}
 $[\rm OII]$ & $21.9^{+2.3}_{-2.3}$   &  $22.0^{+2.1}_{-2.1}$  & $61.3^{+11.1}_{-11.1}$  \\
 \noalign{\smallskip}
  & $1.82^{+0.26}_{-0.30}$ & $0.35^{+0.07}_{-0.07}$ & $0.89^{+0.18}_{-0.19}$ \\
\noalign{\smallskip}
 \noalign{\smallskip}
 \hline
 \noalign{\smallskip}
 \noalign{\smallskip}
 SED & $54.0^{+10.2}_{-10.2}$ & $56.8^{+22.3}_{-13.8} $ & $70.3^{+16.8}_{-13.2}$ \\
 \noalign{\smallskip}
 $^{\rm (\citetalias{weaver_cosmos2020_2022})}$ & $4.49^{+0.95}_{-1.00}$ & $0.90^{+0.38}_{-0.27}$ & $1.02^{+0.26}_{-0.23}$ \\
 \noalign{\smallskip}
 \noalign{\smallskip}
 \noalign{\smallskip}
 \hline
\end{tabular}
 \end{center}
 \tablefoot{For each star-formation indicator, we specify the SFR in the upper row and the relative sSFR in the bottom row, in units of M$_\odot$~yr$^{-1}$ and Gyr$^{-1}$, respectively. Star-formation indicators from top to bottom are: the FIR flux we estimated by integrating the dust component of the SED; the integrated FIR flux but with the values reported by \citet{kartaltepe_multiwavelength_2010}; the H$\alpha$ emission line from DESI spectra; the $[\rm OII]$ emission line from DESI spectra; the UV-optical SED fitting from \citet{weaver_cosmos2020_2022}.}
\label{tab:sfrs}
\end{table}

Aperture-corrected fluxes are converted to luminosities using the spectroscopic redshift of each galaxy. Dust attenuation is incorporated into the SFR estimates through an attenuation term $A_{\rm H\alpha}$. 
For galaxies with high signal-to-noise H$\beta$ measurements, we estimate nebular extinction directly from the Balmer decrement (H$\alpha$/H$\beta$), adopting a Calzetti attenuation law \citep{calzetti_dust_2000}. For \bggii\ where the H$\beta$ measurement is of lower significance and the Balmer decrement yields highly uncertain attenuation values, we adopt an alternative approach based on the infrared-excess–stellar-mass (IRX–$M_\star$) relation by \citet{mclure2018}. Empirical studies have shown that IRX correlates strongly with stellar mass and exhibits only weak redshift evolution out to at least $z \sim 1$, providing nearly redshift-independent attenuation estimates for star-forming galaxies \citep[\eg][]{pannella_goods-herschel_2015, alvarez-marquez_dust_2016}. The inferred infrared excess is converted into an effective H$\alpha$ attenuation assuming a Calzetti attenuation curve.

With these conventions, the H$\alpha$- and [OII]-based SFRs can be expressed as \citep[see, \eg][]{2010MNRAS.405.2594G,2013ApJ...779..137Z,2020MNRAS.493.5987O}:
\begin{equation}
\label{eq:SFR_Ha}
\frac{{\rm SFR}_{\rm H\alpha}}{\rm M_\odot\,{\rm yr}^{-1}} =
5.0 \times 10^{-42}
\frac{L_{\rm H\alpha}}{{\rm erg\,s}^{-1}}
10^{0.4\,A_{\rm H\alpha}}\;,
\end{equation}

\begin{equation}
\label{eq:SFR_OII}
\frac{{\rm SFR}_{\rm [O\,II]}}{\rm M_\odot\,{\rm yr}^{-1}} =
5.0 \times 10^{-42}
\frac{L_{\rm [O\,II]}}{{\rm erg\,s}^{-1}}
10^{0.4\,A_{\rm H\alpha}}\,r_{\rm lines}^{-1}\;,
\end{equation}
where $r_{\rm lines}$ accounts for the relative attenuation between the [OII] and H$\alpha$ lines under the adopted extinction law.

Following \citet{2010MNRAS.405.2594G}, we adopted $r_{\rm lines}=0.5$, and when considering the [OII] doublet, we have taken the total flux of the doublet components and summed their uncertainties in quadrature.
The H$\alpha$ and [OII] doublet fluxes before aperture correction are in the range $F_{\rm H\alpha} \sim 46 - 146$, and $F_{\rm [O~II]} \sim 12 - 88$, respectively, in units of  $10^{-17}$~erg~s$^{-1}$~cm$^{-2}$. The resulting SFRs for \bggi, \bggii, and \bggiii\ are reported in Table \ref{tab:sfrs}, in the third and fourth rows.

\subsubsection{\textit{SED fitting and IR luminosity with LePhare}}\label{sec:LePhaseSEDfitting}

\begin{figure}[h!]
\centering
\includegraphics[width=0.5\textwidth]{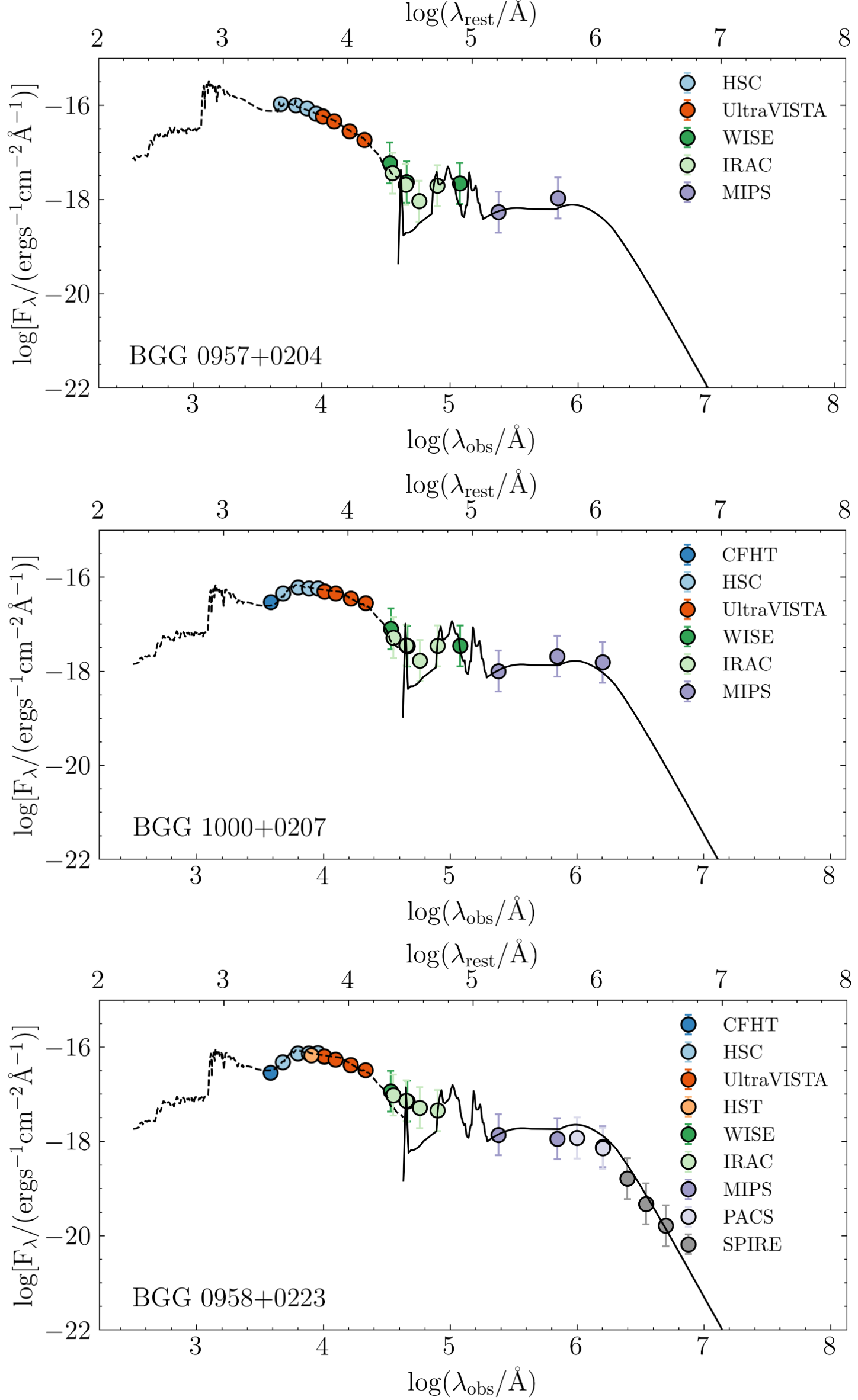}
\caption{SEDs and respective LePhare best fitting modelling for the three target BGGs from $u$-band to FIR. Different colours refer to different instruments, as in the legend (see the text for details about the instruments and relative bands). Dashed and solid lines indicate best fits for the stellar and dust components, respectively.}
\label{fig:SED}
\end{figure}

As an independent estimate of the total star-formation activity, we compute SFRs based on the total infrared luminosities. The far-infrared luminosity ($L_{\rm IR}$), integrating the dust emission over $\sim$8--1000~$\mu$m, provides a direct tracer of obscured star formation that is largely insensitive to the detailed dust geometry affecting optical diagnostics \citep{1998ARA&A..36..189K}.

In order to independently estimate SFRs from infrared luminosities, we performed SED fitting for our three target BGGs.
In line with previously adopted approaches in the literature \citep[\eg][]{castignani_molecular_2019, poitevineau2023}, we performed galaxy SEDs modelling using LePhare \citep{arnouts1999,ilbert_accurate_2006}. The far-infrared photometry was fitted separately in order to capture potential dust-related emission, adopting the set of 105 infrared templates from \citet{charyelbaz2001}. Photometric observations at shorter wavelengths were instead modelled with the \texttt{CE\_NEW\_MOD} template set, which includes a suite of galaxy spectra generated through interpolation.

For the SED fitting, we used photometric archival data from the $u$-band to FIR from the COSMOS2020 galaxy catalogue and from the NASA/IPAC Extragalactic Database (NED). Archival photometry includes optical and near-infrared data from CFHT \citep{sawicki_cfht_2019}, HSC \citep{aihara19}, HST ACS \citep{koekemoer_cosmos_2007} and UltraVISTA \citep{mccracken_ultravista_2012, moneti_vizier_2023}, and infrared from Spitzer/IRAC and MIPS \citep{spitzer2009,kartaltepe_multiwavelength_2010,spitzer2013}, WISE \citep{allwise2014} and Herschel/PACS and SPIRE \citep{pacs_spire_2015}.
Our results for the best-fitting SED modelling using separate handling of stellar and dust components are shown in Fig. \ref{fig:SED} for the three target BGGs.

From the resulting fits, we derived the total infrared luminosity ($L_{\rm IR}$) by integrating the best-fit model of the dust component in the rest-frame 8–1000 $\mu$m range. This luminosity was then translated into an SFR using the \citet{1998ARA&A..36..189K} relation, adjusted for a \citet{Chabrier2003PASP..115..763C} initial mass function:
 \begin{equation}
\label{eq:SFR_IR_K98}
\frac{{\rm SFR}_{\rm FIR}}{\rm M_\odot\,{\rm yr^{-1}}}=
1.075\times10^{-10}\,\left(\frac{L_{\rm IR}}{L_\odot}\right)\,.
\end{equation}

\noindent Dust emission integration yielded the following IR luminosities: $\log(L_{\rm IR}/L_\odot)=10.8$, $11.4$, and $11.6$ for \bggi, \bggii, and \bggiii, respectively, translating into SFR$_{\rm FIR}=(6.7 \pm 1.1)$, $(29.8 \pm 3.6)$, and $(45.8 \pm 5.1)$ M$_\odot$~yr$^{-1}$. We assume these as fiducial values for the rest of the analysis as reported in Table \ref{tab:galaxy_properties}. Based on their high IR luminosities, the three BGGs can be considered luminous infrared galaxies \citep[LIRGs; see \eg][]{sanders_mirabel1996}, which are often associated with significant star formation activity \citep[\eg][]{lirgs2015,lirgs2017,castignani_jablonka_2020}. Although \bggi\ has an IR luminosity slightly below the formal LIRG threshold of $L_{\rm IR}\simeq10^{11}~L_\odot$, it lies very close to this limit and is therefore considered comparable to the rest of the sample.

This conversion of IR luminosity into an SFR estimate is valid provided that the FIR luminosity predominantly arises from reprocessed stellar light associated with young massive stars. We assume this condition holds for our three targets based on the following considerations. 

We performed WISE colour-based classification \citep{stern2012,jarrett2017} using AllWISE photometry \citep{allwise2014}. According to these diagnostics, \bggi\ is classified as an intermediate disk, \bggii\ as a starburst galaxy, and only \bggiii\ as an AGN. This classification is generally consistent with the UV--optical diagnostics reported in the DESI VAC \citep{myers_target-selection_2023}, including the BPT diagnostic \citep{baldwin_classification_1981} whenever available. In particular, \bggi\ is classified as star-forming by all UV--optical diagnostics, while \bggii\ lacks a BPT classification in the DESI catalogue but is identified as star-forming (not AGN) by all the other available methods\footnote{Further details about the methods used for DESI galaxy/AGN classification can be found at \url{https://data.desi.lbl.gov/doc/releases/dr1/vac/agngal}.}. \bggiii\ instead presents a mixed classification, being identified as star-forming according to the BPT diagnostic but as an AGN based on its WISE colours. In line with this, it is worth mentioning that \bggiii\ has WISE colours W1--W2~=~0.86~mag (Vega) and W2--W3~=~2.88~mag (Vega), placing it close to the boundary between the AGN and intermediate-disk classes.

However, for all our targets, the FIR emission is well fitted by the templates of \citet{charyelbaz2001}. This suggests that the FIR emission primarily originates from dust heated by star formation, with minimal AGN contamination, which would instead result in a steep-spectrum SED. The absence of a strong power-law component in the SED of \bggiii\ (see Fig.~\ref{fig:SED}) suggests the assumption of dominant dust emission holds for this target as well.

Finally, we compared our estimates of the total infrared luminosities with the ones inferred by \citet{kartaltepe_multiwavelength_2010}. However, only luminosities for \bggii\ and \bggiii\ are provided in this study. Reported FIR luminosities for \bggii\ and \bggiii\ are $\log(L_{\rm IR}/L_\odot)=11.64\pm0.03$ and $11.45\pm0.04$, respectively, yielding SFR$_{\rm FIR}=(49.6 \pm 3.2)$ and $(30.3 \pm 2.8)$ M$_\odot$~yr$^{-1}$, as reported for comparison in Table \ref{tab:sfrs}. Overall, the resulting SFRs agree well with those derived from our FIR-based estimates.

Table \ref{tab:sfrs} summarises and compares the results from different multiwavelength SFR indicators we considered in this analysis, including the FIR emission we estimated by integrating the dust component of our best-fit SED modelling, the integrated FIR-based SFR inferred from fluxes reported by \citet{kartaltepe_multiwavelength_2010}, the H$\alpha$ and $[\rm OII]$ emission line from DESI spectra, and the UV-optical SED fitting from \citet{weaver_cosmos2020_2022}.

For each BGG, the different SFR estimates are broadly consistent, agreeing within a factor of $\sim3$, which is in line with expectations when comparing different SFR tracers \citep[\eg][]{kennicutt_star_2012, figueira2022}.
However, the comparison of star formation rates derived from multiple diagnostics also highlights some variations among the different tracers, possibly reflecting their sensitivity to distinct physical timescales and levels of dust obscuration. In particular, the larger discrepancy observed for \bggi\ possibly reflects the limitations of UV--optical SED-based estimates in the presence of obscured star formation in dusty systems. In general, UV--optical SED fitting without IR constraints tends to yield higher SFRs in our sample, while optical emission-line indicators provide lower or intermediate values that may be affected by extinction and non-star-forming excitation mechanisms \citep[see \eg][]{kennicutt_star_2012,conroy2013,2020MNRAS.493.5987O}. As they trace the dust emission heated by star formation, infrared-based estimates typically offer a more robust measure of the star formation rates and are therefore adopted as fiducial values in the following analysis. Therefore, we used the estimated values obtained via our FIR-based SED fitting that we performed with LePhare (see Table~\ref{tab:sfrs}). The diversity of SFR estimates underscores the complex and heterogeneous nature of star formation in these systems. A more detailed interpretation of the individual galaxies, and of the possible physical origin of the discrepancies among the various SFR indicators, is presented in Sect.~\ref{sec:discussion}.

\section{Millimetre observations and data reduction}\label{sec:mm_observations}

In this Section, we describe our millimetre observations of the three BGG targeted in this work, and the relative data reduction.

\subsection{IRAM~30m observations}\label{sec:IRAM30m_observations}

We observed the three sources using the Eight Mixer Receiver (EMIR) mounted onto the IRAM~30m telescope at Pico Veleta in Spain. The observations were carried out in July, September, October, and November 2025 (ID: 079-25; PI: G.~Toni \& G. Castignani). Altogether, each source was observed for about 2.3 to 3.3 hours (double polar).

Each of the E090, E150, and E230 receivers, operating between 1 and 3~mm, offers 4$\times$4~GHz instantaneous bandwidth covered by the correlators.
For all three sources we used the E090 and E230 receivers to simultaneously target the CO(1$\rightarrow$0) and CO(3$\rightarrow$2) lines, redshifted to $\sim(3.3-3.5)$~mm and $\sim(1.1-1.2)$~mm, respectively. The wobbler-switching mode was used for all observations to minimise the impact of atmospheric variability.
The Wideband Line Multiple Autocorrelator (WILMA) was used to cover the 4$\times$4~GHz bandwidth in each linear polarisation.   We also simultaneously recorded the data with the fast Fourier Transform Spectrometers (FTS) as a backup at 200 kHz resolution.

\begin{figure}[h!]
\centering
\includegraphics[width=0.5\textwidth]{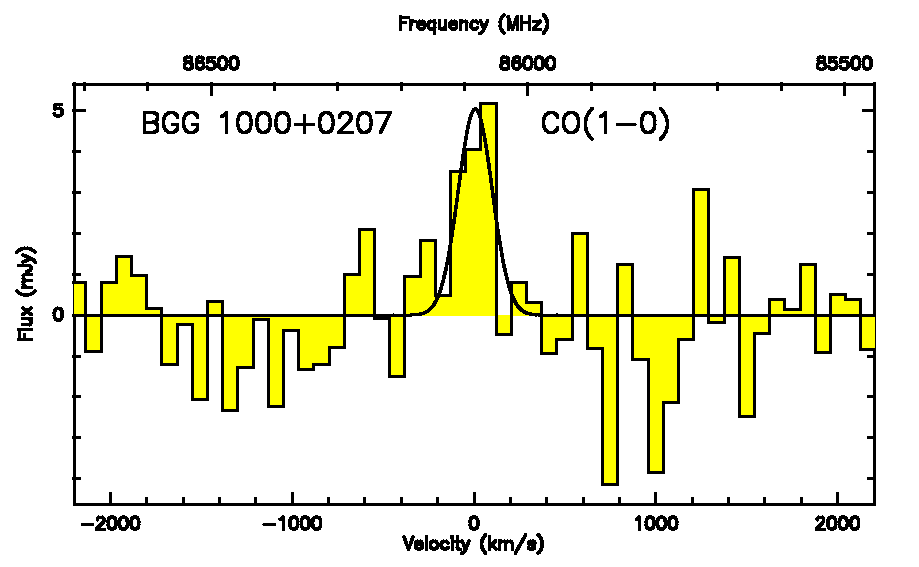}
\caption{Baseline-subtracted  CO(1$\rightarrow$0) spectrum of 
BGG~1000+0207. The solid line shows the Gaussian best ﬁt. The ﬂux (y-axis) in units of mJy is plotted against the relative velocity with respect to the BGG redshift.}
\label{fig:COspectrum_BGG2}
\end{figure}

The IRAM~30m half-power beam width (HPBW) is $\sim16$~arcsec~$\frac{\lambda_{\rm obs}}{2~{\rm mm}}$ \citep{Kramer2013}, where $\lambda_{\rm obs}$ is the observer frame wavelength. This translates into a beam diameter for our three targets falling in the range $\sim (26 - 28)$$^{\prime\prime}$ and $\sim (8.7 - 9.4)$$^{\prime\prime}$ for CO(1$\rightarrow$0) and CO(3$\rightarrow$2), respectively. All sources are thus unresolved by our IRAM~30m observations. Indeed, the typical CO-to-optical spatial extent ratio is $\sim0.5$ \citep{Young1995}, and in general of the order of unity. 

We encountered overall unfavourable weather conditions in July and September, often characterised by atmospheric instability, which is common,  considering that the observations were carried out during the local summer season. However, in late October and late November, we had optimal weather conditions. Altogether, we had system temperatures ranging between $T_{\rm sys}\simeq$~115$-$124~K at $\sim$~3~mm. At $\sim$~1~mm system temperature ranged $T_{\rm sys}\simeq$~343$-$424~K. 

The data reduction and analysis were performed using the {\tt\string CLASS} software of the {\tt\string GILDAS}  package\footnote{\url{https://www.iram.fr/IRAMFR/GILDAS/}}. 
We applied standard efficiency corrections to convert 1) the antenna temperature $T$a$^\ast$ into the main beam temperature $T_{\rm mb}$, and then 2) $T_{\rm mb}$ into the corresponding CO line flux, with a conversion factor of 5~Jy/K. In particular, we adopted the following efficiency  corrections: {$T_{\rm mb}/T{\rm a}^\ast$} = 1.17 and 1.70, at $\sim3$~mm and $\sim1$~mm, respectively.\footnote{\url{https://www.iram.es/IRAMES/mainWiki/Iram30mEfficiencies}}. 

We detected the CO(1$\rightarrow$0) line for BGG~1000+0207. The baseline-subtracted spectrum is shown in Fig. \ref{fig:COspectrum_BGG2}. From the best-fit Gaussian model, we measure a full width at half maximum of $\mathrm{FWHM} = 220 \pm 70$ km/s. This value is consistent, within uncertainties, with the typical line widths observed in massive BCGs at similar redshifts \citep[\eg][]{edge_detection_2001,Castignani_clash_newitem_2020,dunne_co_2021}. Conversely, we were able to set upper limits for the other BGGs analysed in this work. Concerning CO(1$\rightarrow$0) lines of BGG~0957+0204 and 0958+0223, we reached a root mean square (rms) of 1.35 and 1.32~mJy, for velocity channels of 97.75 and 105.7~~km/s, respectively, after removing the baseline with a linear fit. 
Similarly, for the CO(3$\rightarrow$2) line of BGG~0957+0204, 1000+0207, and 0958+0223, we reached rms values of  6.66, 3.95, and 4.57, with velocity channels of 65.58, 69.65, and 70.49, respectively. In the cases of non-detection, we were then able to set 3$\sigma$ upper limits to the velocity integrated flux of $S_{\rm CO(J\rightarrow J-1)}\,\Delta\mathrm{v}$, where we assumed a velocity resolution and width of 300~km/s, which is typical of massive galaxies such as those in our samples, as mentioned before \citep[see \eg][]{edge_detection_2001,dunne_co_2021}.

\subsection{Molecular gas reservoirs}\label{sec:gas_reservoirs}

For our three sources, we inferred measurements and  3$\sigma$ upper limits to the CO(J$\rightarrow$J-1) velocity integrated luminosities, $L^{\prime}_{\rm CO(J\rightarrow J-1)}$, in units {of} K~km~s$^{-1}$~pc$^2$, and the velocity-integrated CO(J$\rightarrow$J-1) fluxes, $S_{\rm CO(J\rightarrow J-1)}\,\Delta \mathrm{v} $, in units {of} Jy~km~s$^{-1}$. To this aim, we used the following relation from \citet{solomon_molecular_2005}:
\begin{equation}
\label{eq:LpCO}
 L^{\prime}_{\rm CO(J\rightarrow J-1)}=3.25\times10^7\,S_{\rm CO(J\rightarrow J-1)}\,\Delta \mathrm{v} \,\nu_{\rm obs}^{-2}\,D_L^2\,(1+z)^{-3}\,,
\end{equation}
where $\nu_{\rm obs}$ is the observed frequency in GHz of the CO(J$\rightarrow$J-1) transition, $D_L$ is the luminosity distance in Mpc, and $z$ is the redshift of the target galaxy. 

We then estimated the total $H_2$ gas mass, $M_{H_2}$, of the BGGs or its upper limit directly from the velocity-integrated CO line luminosities, given that 
$M_{H_2} = \alpha_{\rm CO}\;L^{\prime}_{\rm CO(1\rightarrow0)} = \alpha_{\rm CO}\;L^{\prime}_{\rm CO(J\rightarrow J-1)}/r_{\rm J1}$, where $\alpha_{\rm CO}$ is the CO-to-$H_2$ conversion factor and $r_{\rm J1}=L^{\prime}_{\rm CO(J\rightarrow J-1)}/L^{\prime}_{\rm CO(1\rightarrow0)}$ denotes the excitation ratio. For the two galaxies without clear CO detection, we used CO(1$\rightarrow$0) flux to estimate $M_{H_2}$ upper limits. We adopted a Galactic CO-to-H$_2$ conversion factor 
$X_{\mathrm{CO}} \approx 2 \times 10^{20}\ \mathrm{cm^{-1}\ (K\,km\,s^{-1})^{-1}}$, 
namely, 
$\alpha_{\mathrm{CO}} = 4.36\ \rm M_\odot\ (\mathrm{K\,km\,s^{-1}\,pc^{2}})^{-1}$ \citep{solomon_molecular_1997, bolatto_co--h2_2013}. The results for the molecular gas properties are reported in Table \ref{tab:properties_mol_gas}. The choice of a typical Galactic conversion factor is in line with that adopted in previous studies and thus enabled us to perform a homogeneous comparison, presented in Sect. \ref{sec:comparison_literature} \citep[\eg][]{geach_evolution_2011,castignani_molecular_2019,Castignani_clash_newitem_2020,castignani_molecular_2020,dunne_co_2021}. 
However, our selected galaxies have FIR-based SFR $\sim10-50$~M$_\odot$~yr$^{-1}$, with specific star-formation rates (sSFR)
in the range between $\sim 3.4-6.1\times {\rm sSFR}_{\rm MS}$ (see Table~\ref{tab:galaxy_properties}), hence occupying the regime above the star-forming main sequence, which is typically characterised by sSFR values larger than $3 \times \rm {\rm SFR}_{\rm MS}$. Therefore,  a lower conversion factor, typical of starburst galaxies, $\alpha_{\mathrm{CO}} \sim 0.8 \, \rm M_\odot\ (\mathrm{K\,km\,s^{-1}\,pc^{2}})^{-1}$, could also be an appropriate choice in our specific case. We discuss the effect of adopting a starburst $\alpha_{\mathrm{CO}}$ conversion factor below in Sect. \ref{sec:implications}.

\begin{table*}[tb]
\caption{Molecular gas properties resulting from our CO observations.}
\centering
\begin{center}
\begin{adjustbox}{width=1.0\textwidth, left}
\begin{tabular}{cccccccccccccc}
\hline
\hline\\
  &   &    &   & integration &  &  &  &  &  \\
 Galaxy  &  $z_{spec}$ &   CO(J$\rightarrow$J-1)  &  $\nu_{\rm obs}$ & time & $S_{\rm CO(J\rightarrow J-1)}\,\Delta\varv$   &  $L^\prime_{\rm CO(J\rightarrow J-1)}$ &  $M_{H_2}$ & $\tau_{\rm dep}$ & $\frac{M_{H_2}}{M_\star}$ \\ 
   &  & & [GHz] & [h] &   [Jy~km~s$^{-1}$] & [$10^9$~K~km~s$^{-1}$~pc$^2$] & [$10^{10}~\rm M_\odot$] & [Gyr] & \\ 
 (1) & (2) & (3) & (4) & (5) & (6) & (7) & (8) & (9) & (10)  \\\\
 \hline\\
BGG~0957+0204 & 0.25288 & 1$\rightarrow$0 & 92.012 & 2.28 & $<0.694$ & $<2.21$ & $<0.96$ ($<0.18)$ & $<1.43$ ($<0.26$) & $<0.80$ ($<0.15$)
\\           &  & 3$\rightarrow$2 & 276.023  & 2.28  & $<2.80$ & $<0.99$ \\\\
\hline\\
BGG~1000+0207  & 0.33904 &  1$\rightarrow$0 & 86.085  & 3.33 & $1.18^{+0.31}_{-0.31}$ & $6.90^{+1.81}_{-1.81}$ & $3.01^{+0.79}_{-0.79}$ ($0.55^{+0.15}_{-0.15}$)  &  $1.01^{+0.29}_{-0.29}$ ($0.19^{+0.05}_{-0.05}$) & $0.48^{+0.15}_{-0.15}$  ($0.09^{+0.03}_{-0.03}$) \\
          &  & 3$\rightarrow$2 & 258.242  & 3.33  & $<1.71$ & $<1.11$ \\\\
\hline\\ 
 BGG~0958+0223 & 0.35510 & 1$\rightarrow$0 & 85.065 & 3.25 & $<0.705$ & $<4.54$ & $<1.98$ ($<0.36$)   & $<0.43$ ($<0.08$) & $<0.29$ ($<0.05$)  \\
 &  & 3$\rightarrow$2 & 255.181  & 3.25  & $<1.99$ & $<1.42$ \\\\
\hline

\end{tabular}
\end{adjustbox}
\end{center}
\tablefoot{Column description: (1) galaxy name;  (2) spectroscopic redshift as in Table~\ref{tab:galaxy_properties}; (3-4) CO(J$\rightarrow$J-1) transition and observer frame frequency; (5) on-source integration time (double polar); (6)  CO(J$\rightarrow$J-1) velocity-integrated flux; (7) CO(J$\rightarrow$J-1) velocity-integrated luminosity; (8) molecular gas mass $M_{H_2} =\alpha_{\rm CO} L^\prime_{\rm CO(1\rightarrow0)}$; (9) depletion timescale  $\tau_{\rm dep}=M_{H_2}/{\rm SFR}$; (10) molecular gas-to-stellar mass ratio. For columns (8-10), we assume a conversion factor of $\alpha_{\rm CO}=4.36~\rm M_\odot\,({\rm K~km~s}^{-1}~{\rm pc}^2)^{-1}$; values in parentheses correspond to those obtained assuming a starburst conversion factor, $\alpha_{\rm CO}=0.8~\rm M_\odot\,({\rm K~km~s}^{-1}~{\rm pc}^2)^{-1}$. Columns (6-10) show 3$\sigma$ upper limits, except for the CO(1$\rightarrow$0) detection value for BGG~1000+0207.}
\label{tab:properties_mol_gas}

\end{table*}

\section{Discussion}
\label{sec:discussion}

Given the star-forming nature of these galaxies, as confirmed by their colours and by multiple independent SFR indicators, and considering the presence of disturbed morphologies indicative of interactions, one would generally expect significant CO emission tracing large molecular gas reservoirs \citep[\eg][]{violino_galaxy_2018,lisenfeld_co_2019,castignani_star-forming_2022}. 
However, as presented in Sect. \ref{sec:mm_observations} and summarised in Table~\ref{tab:properties_mol_gas}, two out of the three BGGs are not detected in CO, yielding upper limits on their molecular gas masses being below $\sim 10^{10}$ M$_\odot$. These limits imply molecular-gas-to-stellar-mass ratios in the range of $M_{H_2}/M_\star\lesssim0.3-0.8$. By combining the molecular gas estimates (or their upper limits) with the relatively high SFR values, as those inferred in particular for \bggii\ and \bggiii, we obtain a variety of depletion timescales $\tau_{\rm dep}=M_{H_2}/$SFR, ranging from the shortest value of 400~Myr (for \bggiii) up to $\sim1$~Gyr or possibly above, for \bggi\ and \bggii.

Although all are star-forming galaxies, a possible interpretation is that these BGGs are observed at different phases of their evolution. Below, we discuss how environment-driven processes could contribute to the depletion of gas reservoirs and to the quenching of star formation in these galaxies.

\subsection{\bggi}
\label{sec:discussionBGG1}
\bggi\ displays a disturbed morphology and is likely undergoing interaction with a nearby companion galaxy, which can be clearly seen in Fig.~\ref{fig:images}. Both galaxies are unresolved by our CO observations, within the IRAM~30m beam. Major interactions and mergers are generally expected to be gas-rich systems, as illustrated by nearby interacting galaxies such as the Antennae system and by statistical studies at low and intermediate redshift \citep[\eg][]{2000ApJ...542..120W,saintonge2012}. In this context, the absence of a CO detection in \bggi\ is particularly noteworthy.

The molecular gas mass is constrained from the rms of the CO observations, yielding an upper limit of $M_{H_2} < 9.6 \times 10^{9}$~M$_\odot$. Even accounting for the fact that two galaxies may contribute within the beam, this limit implies an upper limit to the total molecular gas content of $M_{H_2}/M_\star<0.8$. Adopting our FIR-based SFR estimate, the resulting depletion time is $\tau_{\rm dep} \lesssim 1.4$~Gyr, consistent with typical values for main-sequence galaxies at similar redshift. However, shorter values are not excluded, given the upper limit.

The combination of an interaction-driven morphology and the upper limit to the molecular gas content, despite the ongoing star formation activity, suggests an evolving system. Rather than being driven solely by interaction-induced star formation, the lack of detectable molecular gas likely points to environmentally driven processes, such as starvation due to suppressed gas accretion within the group environment. In this scenario, the interaction may have accelerated gas consumption, while the surrounding environment could hinder the replenishment of the cold gas reservoir, which would be consistent with the observed non-detection in CO.

To explain the observed phenomenology, another interpretation may be that the galaxy experienced a recent episode of enhanced star formation.  The IR-based SFR estimate, $\sim 7$~M$_\odot$~yr$^{-1}$, is indeed smaller than line-based SFR estimates ($H\alpha$, [OII]), in the range of $\sim$13--22~M$_\odot$~yr$^{-1}$. This difference, given that the former estimate generally probes star formation on a larger timescale ($\sim100$~Myr) than the latter ($\sim10$~Myr), suggests that this BGG may have recently experienced an enhancement of the star formation activity, possibly induced by galaxy-galaxy interaction and gas infall, followed by the consumption of the molecular gas.

\begin{figure*}[h!]
\centering
\includegraphics[width=\textwidth]{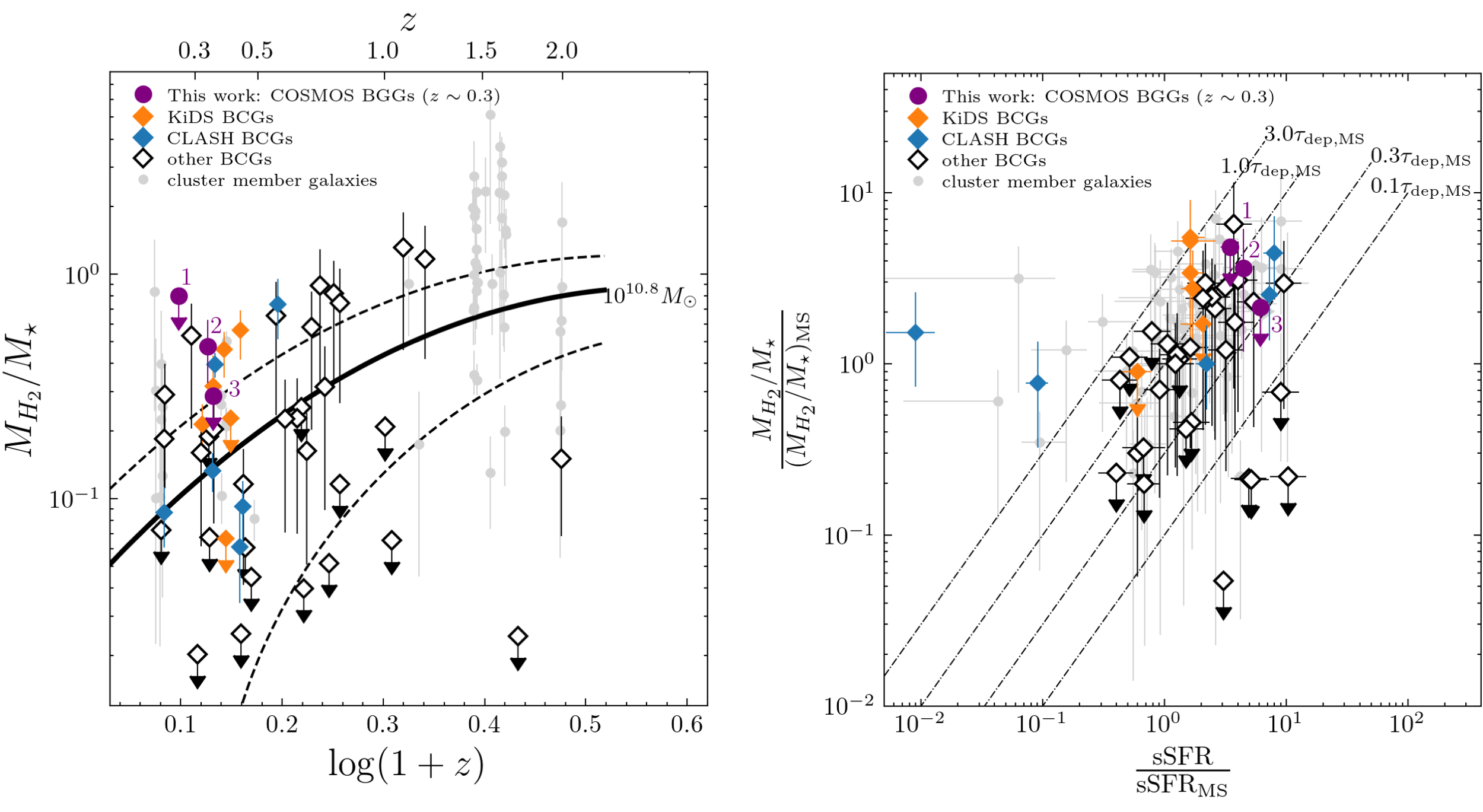}
\caption{Molecular gas properties of BCGs (diamonds) and cluster galaxies (grey points) detected in CO from the literature, compared to the BGGs studied in this work (filled purple circles: 1. \bggi; 2. \bggii; 3. \bggiii). BCGs marked in orange are KiDS BCGs \citep{castignani_star-forming_2022,castignani_star-forming_2023} and those marked in blue are CLASH BCGs \citep{Castignani_clash_newitem_2020,fogarty_dust_2019}. Left: Molecular gas--to--stellar mass ratio as a function of redshift. The solid line shows the main-sequence scaling relation for field galaxies with $\log(M_\star/\rm M_\odot)=10.8$ from \citet{tacconi_phibss_2018}, with dashed lines indicating the $1\sigma$ uncertainty. Right: Molecular gas fraction versus specific star formation rate, both normalised to their main-sequence values following \citet{tacconi_phibss_2018} and \citet{speagle_highly_2014}. Dot--dashed lines indicate constant molecular gas depletion times in units of the main-sequence depletion time. All values are calibrated to the same $\alpha_{\rm CO}$ conversion factor for a consistent comparison.
}
\label{fig:irampost}
\end{figure*}

\subsection{\bggii}
\bggii\ exhibits a red central component surrounded by blue substructures or satellite-like features and an elongated structure towards north-west (see Fig.~\ref{fig:images}). These features indicate a complex morphology and, possibly, spatially distinct stellar populations. The BGG is also gas rich, given our detection of a substantial molecular gas reservoir, $M_{H_2} \simeq 3 \times 10^{10}$~M$_\odot$, yielding a 
large molecular-gas-to-stellar-mass ratio of $M_{H_2}/M_\star\simeq0.5$, even higher than the values of $M_{H_2}/M_\star\simeq0.1-0.4$, previously found in highly star forming (SFR $\gtrsim100~$M$_\odot$~yr$^{-1}$) intermediate-redshift BCGs such as RX1532 \citep{Castignani_clash_newitem_2020} and M1932 \citep{fogarty_dust_2019} from the CLASH survey \citep{Postman2012}. Such a large molecular gas reservoir for \bggii, together with a substantial IR-based SFR  of $\sim30$~M$_\odot$~yr$^{-1}$, yields a depletion time $\tau_{\rm dep} \simeq1.0$~Gyr, as typically found in main sequence star-forming galaxies. Altogether, these features suggest that the gas fuelling and sustaining of the ongoing star formation is effective, while the molecular gas reservoir is also being consumed.

However, our IR-based SFR estimate of $\sim30$~M$_\odot$~yr$^{-1}$ is higher than the line-based SFR values  inferred from H$\alpha$ and [OII], in the range $\sim(12-22)$~M$_\odot$~yr$^{-1}$. Based on the different timescales probed by these two estimators, these values suggest a possible decline of the star formation activity over the past $\sim100$~Myr. Therefore, \bggii\ may be observed in a phase preceding the gas exhaustion and the onset of quenching.

Another interesting aspect of \bggii\ concerns its molecular gas excitation. We quantify the molecular gas excitation ratio as the ratio. For \bggii, we derive an upper limit of $r_{31} < 0.16$. This value is significantly low compared to typical values observed in star-forming galaxies, starburst or AGN hosts, where $r_{31} \sim 0.4-$0.8 is commonly found, and even lower than those measured in many BCGs \citep[\eg][]{Bauermeister2013,tacconi_phibss_2018,Lamperti2020,Castignani_clash_newitem_2020}. An explanation for this unusually small excitation ratio might be that the average molecular gas volume density is likely insufficient to significantly populate the $J=3$ level, resulting in weak or undetectable CO(3$\rightarrow$2) emission. 
In addition to this scenario, observational effects may further contribute to suppressing the ratio due to the wavelength dependence of the telescope beam size. At $\lambda_{\rm obs} \sim 1$mm, corresponding to the redshifted CO(3$\rightarrow$2) line, the beam probes primarily the central region of the system, whereas at $\lambda_{\rm obs} \sim 3$mm, corresponding to the redshifted CO(1$\rightarrow$0) line, the larger beam encompasses the more extended molecular structure, including the blue substructures and satellites (see Fig.~\ref{fig:images}). If a fraction of the molecular gas resides in these extended components, the CO(1$\rightarrow$0) emission would be enhanced relative to CO(3$\rightarrow$2), lowering $r_{31}$.
The observed upper limit may therefore reflect a combination of intrinsically low gas densities and differential beam sampling, both consistent with an extended, dynamically disturbed molecular gas reservoir.

\subsection{\bggiii}

\bggiii\ has a disturbed morphology, suggesting recent or ongoing dynamical perturbations. It also seems to host an AGN, as indicated by its optical emission-line properties and as confirmed by the WISE colour classification described in Sect. \ref{sec:LePhaseSEDfitting}. This galaxy is the only BGG in our sample to exhibit a broad H$\alpha$ emission-line component in the DESI spectrum\footnote{As previously mentioned in Sect.~\ref{sec:SFR_DESIspectra}, we stress that the H$\alpha$ emission line
was decomposed into narrow and broad emission line components, and only the former was used to estimate the H$\alpha$-based SFR.}, characterised by a line width of $\sigma=2300$~km/s and a monochromatic luminosity of $1.23\times10^{42}$~erg/s, as typically found in the broad line region of Type~1 AGN \citep[\eg][]{stern_laor_2012,liu_BLR_2019}. Interestingly, these values yield a single-epoch black hole mass of $\log(M_{\rm BH}/M_\odot) \simeq 8.26$, obtained using the calibration provided by \citet{Shen2011}. High black hole masses,  $\log(M_{\rm BH}/M_\odot) \gtrsim 8$, are indeed commonly found in giant ellipticals, as those hosted by cluster central galaxies and BCGs in particular \citep[e.g.][]{Lauer2007,Phipps2019}. \bggiii\ is also the only one among our targets to have archival X-ray data, from the XMM-{\it Newton} telescope \citep[][]{cappelluti09}, which yield an X-ray luminosity of $L_{\rm 2-10\,keV} \simeq 3.7 \times 10^{43}$ erg s$^{-1}$. Given the estimated black hole mass and assuming a fiducial bolometric correction, $k_{\rm bol} = L_{\rm bol}/L_{\rm 2-10\,keV} \simeq 20$, typical of quasars \citep[\eg][]{Lusso2012,Duras2020}, the corresponding Eddington ratio is $\lambda_{\rm Edd} \simeq 0.03$ $(k_{\rm bol}/20)$, where we explicitly retain the dependence on the assumed bolometric correction. This corresponds to an accretion rate within the range of 0.1$-$0.5~M$_\odot$~yr$^{-1}$ that is typically observed in Type~1 X-ray AGN \citep[e.g.][]{Trump2009,Lusso2012}.
Despite the evidence of AGN emission, the IR SED lacks a strong power-law component, indicating that the AGN does not dominate the far-infrared emission (see Fig.~\ref{fig:SED}). Moreover, the overall agreement among SFR estimates from different tracers (Table~\ref{tab:sfrs}) further strengthens the reliability of these estimates.

Concerning the molecular gas properties, our CO observations yield an upper limit of $M_{H_2} < 1.98 \times 10^{10}$~M$_\odot$, corresponding to a relatively short depletion time of $\sim$400~Myr (given SFR $\simeq46$~M$_\odot$~yr$^{-1}$), which is approximately a factor of three lower than the typical depletion time for a main-sequence galaxy with similar properties (see Sect. \ref{sec:comparison_literature} for further details).
Furthermore, \bggiii\ has an IR-based SFR that is smaller than that derived from H$\alpha$ and [OII] emission lines (see Table~\ref{tab:sfrs}), similarly to \bggi\ (see Sect~\ref{sec:discussionBGG1}). This may suggest a recent enhancement of the star formation activity, on time scales of $\sim$10~Myr, as probed by the line-based SFR estimators.

Taken together, these results may point to an efficient but potentially short-lived episode of star formation, where its fuel, the molecular gas reservoir, is being consumed relatively rapidly. 
\bggiii\ may therefore be observed during a transient phase of enhanced star formation, possibly triggered by galaxy interactions, while AGN feedback has not yet led to the suppression of star formation.

Interestingly, the overall morphology of \bggiii~ recalls the cold gas accretion scenario recently reported by \citet{Zhuang2025} for a BGG undergoing a gas-rich minor merger observed with JWST. In that system, tidal stripping of a low-mass, gas-rich satellite deposits cold gas and dust into the BGG, producing prominent filamentary structures on $\sim$10 kpc scales. The authors further speculate that such an event may lead to a partial rejuvenation of the central galaxy through the injection of new cold gas. Although \bggiii\ appears overall elliptical in morphology, it seems connected to the east–west oriented tidal feature, and the main stellar body shows hints of spiral-like structure (especially in the {\it HST} image; bottom right panel of Fig.~\ref{fig:images}) similar to the filamentary configuration reported by \citet{Zhuang2025}, but perhaps viewed from a different projection angle. While we only derive upper limits on the molecular gas content for this galaxy, deeper observations might lead to better constraints on the molecular gas properties or to a detection. The morphological resemblance to the cold-gas accretion scenario may suggest that \bggiii\ has also been experiencing—or may have recently experienced—an external gas accretion event. Such a configuration would be consistent with a transient phase of enhanced star formation and morphological disturbance, potentially even reflecting a mild rejuvenation episode in an otherwise massive and early-type galaxy with an old stellar population.

\subsection{Molecular gas consumption and environment}\label{sec:implications}

Interestingly, our three target galaxies appear to sample different phases of star formation and molecular gas consumption within a group environment. Despite the diversity of morphologies and star formation properties, the molecular gas constraints play a key role in shaping the interpretation of their evolutionary state.
In all cases, the CO observations provide meaningful constraints on the molecular gas mass and possibly their star formation properties.

As previously discussed, \bggii\ is the only source detected in CO. It hosts a substantial molecular gas reservoir that is consistent with sustaining its ongoing star formation activity, which appears to be above typical main-sequence levels. In contrast, despite having similar star formation rates, the other two BGGs have not been detected in CO. This may suggest that in these galaxies, gas depletion could precede star formation decline in these systems, in agreement with scenarios proposed in previous studies of massive galaxies \citep[\eg][]{Castignani2022_VirgoFil}.

Altogether, the comparison between \bggii\ and the two non-detected systems indicates that, once the molecular gas reservoir starts to be consumed or its replenishment is suppressed, the star-formation decline occurs on different timescales, depending on the specific galaxy properties. As \bggii\ has an IR-based SFR that is larger than line-based SFR, it is possible that \bggii\ has been experiencing a star formation decline over the past 100~Myr, which will possibly lead to quenching, unless a gas replenishment occurs.

Additionally, in this work, we adopted a Galactic CO-to-H$_2$ conversion factor, which is typical of MS galaxies. However, our targets have inferred SFRs above the main sequence (see Table~\ref{tab:sfrs}). Therefore, we consider the possibility of adopting instead a lower conversion factor, more appropriate for starbursts, $\alpha_{\mathrm{CO}} \sim 0.8 \, \rm M_\odot\ (\mathrm{K\,km\,s^{-1}\,pc^{2}})^{-1}$ \citep{bolatto_co--h2_2013}. For comparison, we have reported molecular gas properties for both cases in Table  \ref{tab:properties_mol_gas}. Overall, assuming a starburst conversion factor would reduce $M_{H_2}$ by more than a factor of five, and, consequently, also $\tau_{\mathrm{dep}}$ and $\frac{M_{H_2}}{M_{\star}}$, highlighting the systematic dependence of these quantities on the adopted conversion factor. Although this does not change the qualitative interpretation of the results, it would systematically yield shorter depletion times for our targets.

While interaction-driven star formation may contribute to accelerating gas consumption, the environment may also play a role by limiting the replenishment of the cold gas supply. In such a scenario, starvation caused by the suppression of gas accretion, combined with environmental processing such as tidal interactions or stripping, may accelerate the transition towards lower star formation activity.
Visual inspection of Fig.~\ref{fig:images} suggests a complex morphology for our BGGs, elongated structures, blue (possibly star-forming) components, and nearby companions. As an example, the southern companion of \bggi\ appears elongated along the north-east to south-west direction, with possible tidal features.

Several cases of gas-poor or rapidly evolving systems have been discussed in previous studies \citep[\eg][]{castignani_molecular_2019, castignani_environmental_2020}, particularly in relation to galaxy compactness and clumpy morphologies. In this context, the galaxies presented in this work are consistent with scenarios in which environmental effects may influence the balance between gas fuelling and consumption, possibly resulting in shorter star formation timescales even in systems that show signatures of recent or ongoing interactions.

\subsection{Comparison with the literature}\label{sec:comparison_literature}

The molecular gas properties derived for our sample of BGGs can be further interpreted by comparing them with existing CO observations of massive galaxies in denser environments, in particular BCGs and cluster galaxies in general. Comparing our results with these studies allows us to investigate whether central galaxies in groups follow similar trends to their cluster counterparts, or whether the group environment leads to distinct molecular gas exhaustion. 

Our comparison sample includes 103 galaxies, from 61 different clusters in the interval $0.2 \leq z \leq 2.0$. This sample is extracted from the compilation described in Tables A.1 and A.2 of \citet{castignani_molecular_2020}, further updated with more recent observations in \citet{castignani_star-forming_2023}.

Figure~\ref{fig:irampost} compares the molecular gas properties of the BGGs studied in this work with those collected from the literature. Cluster galaxies are shown as grey points, while BCGs are indicated by diamonds, and our BGGs by filled purple circles. Orange diamonds highlight KiDS BCGs \citep{castignani_star-forming_2022,castignani_star-forming_2023} while blue ones mark CLASH BCGs, including the gas-rich and highly star-forming RX1532 \citep{Castignani_clash_newitem_2020} and M1932 \citep{fogarty_dust_2019}, located in cool-core clusters at similar redshifts to our targets.  In the left-hand panel, the molecular gas--to--stellar mass ratio is shown as a function of redshift. The solid line represents the scaling relation for main-sequence field galaxies derived by \citet{tacconi_phibss_2018}, calculated at $\log(M_\star/\rm M_\odot)=10.8$, which corresponds to the median stellar mass of our sample, while the dashed lines indicate the associated $1\sigma$ intervals. This comparison highlights how the BGGs studied in this work span a range of gas fractions that overlaps with, or is slightly higher than, those of gas-rich cluster galaxies at similar redshifts.

The right-hand panel shows the molecular gas fraction as a function of the specific star formation rate, both normalised to their respective main-sequence values following the prescriptions of \citet{tacconi_phibss_2018} and \citet{speagle_highly_2014}. The dot--dashed lines indicate loci of constant molecular gas depletion time in units of the main-sequence depletion time. In this plane, our target BGGs occupy an intermediate region between extreme, cooling-flow–fed BCGs with intense star formation, such as RX1532 and M1932 in the CLASH sample, and the less star-forming and less gas-rich BCGs identified in the KiDS survey and hosted by galaxy clusters.

Specifically, the BGGs studied here seem to exhibit depletion times that are comparable or shorter than those of typical star-forming cluster galaxies (from KiDS and CLASH samples). However, our BGGs also show less extreme star-formation activity than that observed in strongly star-bursting, cooling-flow systems. This is consistent with a scenario in which central galaxies in the group regime may experience efficient gas consumption without the continuous replenishment associated with cluster-scale cooling flows. In this picture, star formation may be sustained only transiently and may be regulated by a combination of limited gas supply and environmental processing, possibly suggesting that BGGs occupy an intermediate regime between gas-rich, highly star-forming cluster core galaxies and more quenched central galaxies in massive haloes, typical of cluster environments.

\section{Conclusions}
\label{sec:conclusions} 
We presented IRAM 30m CO observations of three BGGs extracted from the AMICO-COSMOS group sample \citep{toni_amico-cosmos_2024}, aimed at constraining their molecular gas content and the efficiency of star formation in group environments. The three systems exhibit diverse morphologies, including disturbed structures, blue subcomponents, and signatures of interactions, indicative of ongoing or recent dynamical activity.

Among the targets, only \bggii\ is detected in CO, revealing a substantial molecular gas reservoir of $M_{H_2} \sim 3 \times 10^{10}$~M$_\odot$, while \bggi\ and \bggiii\ remain undetected, with upper limits of $M_{H_2} \lesssim 2 \times 10^{10}$~M$_\odot$. When combined with multiwavelength SFR estimates, these measurements point to distinct star formation regimes: \bggii\ is characterised by sustained star formation with a depletion time of order $\sim1$~Gyr. On the basis of star formation diagnostic and its disturbed morphology, \bggi\ shows some evidence in favour of a recent episode of enhanced star formation, possibly followed by gas exhaustion. \bggiii\ exhibits a relatively short depletion time ($\sim0.3 \, \tau_{dep,\mathrm{MS}}$), indicative of a transient and efficient episode of star formation, possibly followed by gas exhaustion.

Altogether, our results indicate that intermediate-$z$ star-forming BGGs in group environments may occupy an intermediate position, in terms of their molecular gas content and star formation activity, between strongly star-forming, cooling-flow–fed BCGs in massive clusters and less star-forming BCGs. While the molecular gas reservoirs and star formation activity of the BGGs in this work are less extreme than those observed in classical cooling-flow systems, they are nonetheless sufficient to sustain short-lived phases of enhanced star formation. This suggests that environmental processes in groups can regulate the availability and consumption of cold gas, driving rapid evolutionary phases even in massive central galaxies.

As an IRAM pilot project, we plan to expand this study by enlarging our sample both within the COSMOS field and in other extragalactic surveys, an effort that will be facilitated by the steadily increasing spectroscopic coverage. A larger statistical sample will allow us to investigate possible dependencies between star formation activity and molecular gas excitation, and to assess whether the trends suggested by our pilot study persist across different evolutionary stages and environments. In this context, the new public catalogue of galaxy groups from the COSMOS-Web survey, extending out to $z = 3.7$ \citep{toni_cosmos-web_2025}, provides a unique opportunity to probe the molecular gas properties of central galaxies at $z>1$, where current CO observational coverage remains sparse (see Fig. \ref{fig:irampost}).

Finally, confirming and disentangling the relative impact of different quenching mechanisms will require high-resolution interferometric observations—such as those achievable with ALMA or NOEMA—to spatially resolve the molecular gas distribution and excitation conditions in central group galaxies \citep[e.g.][]{castignani_molecular_2025}. Such observations will be crucial to link the global gas properties to the underlying physical processes driving star formation and quenching in dense environments.

\begin{acknowledgements}
 We thank the anonymous referee for the useful comments, which helped improve this manuscript. This work is based on observations with the IRAM~30m telescope, project number 079-25 (PI: G. Toni \& G. Castignani). IRAM is supported by INSU/CNRS (France), MPG (Germany) and IGN (Spain). Part of this research uses data obtained with the Dark Energy Spectroscopic Instrument (DESI). DESI construction and operations is managed by the Lawrence Berkeley National Laboratory. To this end, we acknowledge the use of services/data provided by the Astro Data Lab at NSF’s NOIRLab. NOIRLab is operated by the Association of Universities for Research in Astronomy (AURA), Inc. under a cooperative agreement with the National Science Foundation. We thank the extragalactic community based in Bologna, in particular, Salvatore Quai, Francesca Pozzi, Sirio Belli, Andrea Enia, Lucia Pozzetti, Ivan Delvecchio, Livia Vallini, and Meriem Behiri for useful discussions. We thank Alberto Traina for cross-matching ALMA data to exclude existing correspondence with our targets. We acknowledge the use of GAZPAR tools, operated by CeSAM-LAM and IAP, to double-check the results obtained from our LePhare fit. In particular, we thank Olivier Ilbert for very useful suggestions about the SED fitting procedure. For part of this research, GT was supported by the Angelo Della Riccia Foundation Fellowship Grant 2026. LM acknowledges the financial contribution from the 
PRIN-MUR 2022 20227RNLY3 grant “The concordance cosmological model: stress-tests with galaxy clusters” supported by Next Generation EU and from the grant ASI n. 2024-10-HH.0 “Attività scientifiche per la missione Euclid – fase E”. 
 
\end{acknowledgements}
\bibliographystyle{aa}
\bibliography{references_merged_macros}

\end{document}